\documentclass[lettersize,journal]{IEEEtran}
\IEEEoverridecommandlockouts
\usepackage{tex/packages}
\usepackage{siunitx}
\begin{document}
\begin{acronym}
    \acro{CBADC}{control-bounded analog-to-digital converter}
    \acro{CTSDM}[CT-$\Sigma\Delta$M]{continuous-time sigma-delta modulator}
    \acro{CTSD}{continuous-time sigma-delta}
    \acro{DTSDM}{discrete-time sigma-delta modulator}
    \acro{CRFB}{cascade of resonators with feedback}
    \acro{CRFF}{cascade of resonators with feedforward}

    \acro{ENOB}{effective number of bits}
    \acro{SNR}{signal-to-noise ratio}
    \acro{SDR}{signal-to-distortion ratio}
    \acro{SNDR}{signal-to-noise-and-distortion ratio}
    \acro{SQNR}{signal-to-quantization-noise ratio}
    \acro{SC}{switched-capacitor}
    \acro{CT}{continuous-time}
    \acro{DT}{discrete-time}
    \acro{PSD}{power spectral density}

    \acro{NTF}{noise transfer function}
    \acro{STF}{signal transfer function}

    \acro{CI}{chain-of-integrators}
    \acro{LF}{leapfrog}
    \acro{DC}{digital control}
    \acro{AS}{analog system}
    \acro{DE}{digital estimator}

    \acro{RZ}{return-to-zero}
    \acro{NRZ}{non-return-to-zero}
    \acro{SCR}{switched-capacitor-resistor}
    \acro{DAC}{digital-to-analog converter}
    \acro{ADC}{analog-to-digital converter}

    \acro{FIR}{finite impulse response}
    \acro{FS}{full scale}
    \acro{OSR}{oversampling ratio}
    \acro{OTA}{operational transconductance amplifier}
    \acro{PLL}{phase-locked loop}
    \acro{VCO}{voltage-controlled oscillator}
    \acro{PM}{phase-modulated}
    \acro{FM}{frequency-modulated}
    
    \acro{GBWP}{gain-bandwidth product}

    \acro{LMS}{least mean squares}
    \acro{RLS}{recursive least squares}
    \acro{LMMSE}{linear minimum mean square error}
    \acro{SFDR}{spurious-free dynamic range}
    \acro{dBFS}{decibels relative to full scale}
\end{acronym}

\title{Calibrating Control-Bounded ADCs}

\author{Hampus Malmberg, Till Mettler, Thomas Burger, Fredrik Feyling, and Hans-Andrea Loeliger

\thanks{

Hampus Malmberg, Till Mettler, Thomas Burger, and Hans-Andrea Loeliger are with the Department of Information Technology \& Electrical Engineering, ETH Zürich, 8092 Zürich, Switzerland (email: malmberg@isi.ee.ethz.ch, mettleti@student.ethz.ch, burger@iis.ee.ethz.ch, and loeliger@isi.ee.ethz.ch)

Fredrik Feyling is with Department of Electronic Systems, Norwegian University of Science and Technology, 7491 Trondheim, Norway (email: fredrik.e.feyling@ntnu.no)

}}

% \IEEEpubid{This work has been submitted to the IEEE for possible publication. Copyright may be transferred without notice, after which this version may no longer be accessible.}

\maketitle

\begin{abstract}
The paper considers the calibration of control-bounded analog-to-digital converters. 
It is demonstrated that variations of the analog frontend can be addressed by calibrating the digital estimation filter. 
In simulations (both behavioral and transistor level) of a leapfrog analog frontend, the proposed calibration method restores essentially the nominal performance.
Moreover, with digital-filter calibration in mind, 
the paper reformulates the design problem of control-bounded converters and thereby clarifies the role of sampling, desired filter shape, and nominal conversion error.
\end{abstract}


\section{Introduction}
Stabilizing an analog system using a digital control amounts to implicit analog-to-digital conversion.
This is the core idea behind the \ac{CBADC} concept, first proposed in
\cite{LBWB:11,LW:15} and further developed in \cite{M:20,MWL:21}. The \ac{CBADC} perspective allows more general analog frontends, i.e.,
analog system and digital control combinations, at the expense of a post-processing digital estimation step. 
Specifically, it falls on the digital estimator to consolidate the joint effort of the digital control into an estimate
of the input signal fed into the analog system.
A \ac{CBADC} analog frontend partially resembles \acp{CTSDM} \cite{Schreier:2017,R:2011}. However, 
in contrast to \acp{CTSDM}, the analog frontend of a \ac{CBADC} may contain high-order filters 
while having a guaranteed stable nominal operation. 
Moreover, the high-level design of a \ac{CBADC} is a continuous-time filter design task, whereas 
\acp{CTSDM} are typically considered a discrete-time design concept converted into the continuous-time domain \cite{OG:2006}.

Crucially, the digital estimator's ability to estimate relies on its knowledge of the actual analog frontend. 
As reported in \cite{FMWLY:2022,FMWLY:2023}, this makes \acp{CBADC} particularly sensitive to component variations if not accounted for.
In this paper, we address this issue by demonstrating that the digital estimation filter can be calibrated to compensate for such variations.

\section{The Leapfrog Frontend}\label{sec:leap_frog}
\begin{figure*}[tbp]
    \begin{center}
        \resizebox{0.9\textwidth}{!}{
        \begin{circuitikz}[european voltages]

    \node[op amp] (amp1) at (0,0) {$A_1$};
    \node[rground] at (amp1.+) {};
    \node[inner sep=0] (vgnd1) at ($(amp1.-) + (-0.5,-0.5)$) {};
    \draw (vgnd1) to[short, *-] (amp1.-);
    \draw (vgnd1) to[short, *-] ++(0,1.25) to[C, l=$C_1$] ++ (2.875,0) -| (amp1.out);

    \node[op amp, yscale=-1, rotate=180] (quantizer1) at ($(amp1) + (0, -1.5)$) {};
    \draw ($(quantizer1) + (0.5, -0.25)$) -- ++(-0.25,0) -- ++(0, 0.5) -- ++(-0.25, 0);
    \draw ($(quantizer1) + (0.1875, 0)$) -- ++(0.125,0);
    \node (fs) at ($(quantizer1) + (1.5, 0.675)$) {};
    \draw ($(fs) + (-0.125,-0.125)$) 
        -- ++(0.125, 0) 
        -- ++(0,0.25) 
        -- ++(0.125,0) 
        -- ++(0, -0.25) 
        -- ++(0.125, 0)
        -- ++(0, 0.25)
        -- ++(0.125, 0)
        -- ++ (0, -0.25)
        -- ++ (0.125, 0)
        -- ++ (0, 0.25)
        -- ++ (0.125, 0)
        -- ++ (0, -0.25)
        -- ++ (0.125,0 )
    ;
    \draw[{Stealth[length=0.875mm]}-{Stealth[length=0.875mm]}] ($(fs) + (-0.125, 0.25)$) -- node[above] {\tiny $T$} ++(0.25,0); 
    \draw[Arrow] (fs) -- ++(-0.85, 0);
    \node[above] at ($(quantizer1) + (1, 0.675)$) {\tiny CLK};

    \node[rground] at (quantizer1.+) {};

    \draw (amp1.out) to[short, *-] (quantizer1.-);
    \draw (quantizer1.out) -- ++ (-0.5,0) to[R, l=$R_{\kappa_1}$] (vgnd1);

    \draw (vgnd1) -- ++(-0.5,0.5) to[R, l=$R_{\alpha_1}$] ++(0,1.75) node[inner sep=0] (alpha1) {};

    \draw (quantizer1.-) to[open, v^=$x_1(t)$] (quantizer1.+);
    \draw (quantizer1.out) to[open, v_=$s_1(t)$] ($(quantizer1.+) + (0,-0.785)$);

    \node[op amp] (amp2) at ($(amp1) + (5,0)$) {$A_2$};
    \node[rground] at (amp2.+) {};
    \node[inner sep=0] (vgnd2) at ($(amp2.-) + (-0.5,-0.5)$) {};
    \draw (vgnd2) to[short, *-] (amp2.-);
    \draw (vgnd2) to[short, *-] ++(0,1.25) to[C, l=$C_2$] ++ (2.875,0) -| (amp2.out);

    \node[op amp, yscale=-1, rotate=180] (quantizer2) at ($(amp2) + (0, -1.5)$) {};
    \draw ($(quantizer2) + (0.5, -0.25)$) -- ++(-0.25,0) -- ++(0, 0.5) -- ++(-0.25, 0);
    \draw ($(quantizer2) + (0.1875, 0)$) -- ++(0.125,0);
    \node (fs) at ($(quantizer2) + (1.5, 0.675)$) {};
    \draw ($(fs) + (-0.125,-0.125)$) 
        -- ++(0.125, 0) 
        -- ++(0,0.25) 
        -- ++(0.125,0) 
        -- ++(0, -0.25) 
        -- ++(0.125, 0)
        -- ++(0, 0.25)
        -- ++(0.125, 0)
        -- ++ (0, -0.25)
        -- ++ (0.125, 0)
        -- ++ (0, 0.25)
        -- ++ (0.125, 0)
        -- ++ (0, -0.25)
        -- ++ (0.125,0 )
    ;
    \draw[{Stealth[length=0.875mm]}-{Stealth[length=0.875mm]}] ($(fs) + (-0.125, 0.25)$) -- node[above] {\tiny $T$} ++(0.25,0); 
    \draw[Arrow] (fs) -- ++(-0.85, 0);
    \node[above] at ($(quantizer2) + (1, 0.675)$) {\tiny CLK};
    \node[rground] at (quantizer2.+) {};

    \draw (amp2.out) to[short, *-] (quantizer2.-);
    \draw (quantizer2.out) -- ++ (-0.5,0) to[R, l=$R_{\kappa_2}$] (vgnd2);

    \draw (vgnd2) -- ++(-0.5,0.5) to[R, l=$R_{\alpha_2}$] ++(0,1.75) node[inner sep=0] (alpha2) {};

    \draw (quantizer2.-) to[open, v^=$x_2(t)$] (quantizer2.+);
    \draw (quantizer2.out) to[open, v_=$s_2(t)$] ($(quantizer2.+) + (0,-0.785)$);

    \node[op amp] (amp3) at ($(amp2) + (5,0)$) {$A_3$};
    \node[rground] at (amp3.+) {};
    \node[inner sep=0] (vgnd3) at ($(amp3.-) + (-0.5,-0.5)$) {};
    \draw (vgnd3) to[short, *-] (amp3.-);
    \draw (vgnd3) to[short, *-] ++(0,1.25) to[C, l=$C_3$] ++ (2.875,0) -| (amp3.out);

    \node[op amp, yscale=-1, rotate=180] (quantizer3) at ($(amp3) + (0, -1.5)$) {};
    \draw ($(quantizer3) + (0.5, -0.25)$) -- ++(-0.25,0) -- ++(0, 0.5) -- ++(-0.25, 0);
    \draw ($(quantizer3) + (0.1875, 0)$) -- ++(0.125,0);
    \node (fs) at ($(quantizer3) + (1.5, 0.675)$) {};
    \draw ($(fs) + (-0.125,-0.125)$) 
        -- ++(0.125, 0) 
        -- ++(0,0.25) 
        -- ++(0.125,0) 
        -- ++(0, -0.25) 
        -- ++(0.125, 0)
        -- ++(0, 0.25)
        -- ++(0.125, 0)
        -- ++ (0, -0.25)
        -- ++ (0.125, 0)
        -- ++ (0, 0.25)
        -- ++ (0.125, 0)
        -- ++ (0, -0.25)
        -- ++ (0.125,0 )
    ;
    \draw[{Stealth[length=0.875mm]}-{Stealth[length=0.875mm]}] ($(fs) + (-0.125, 0.25)$) -- node[above] {\tiny $T$} ++(0.25,0); 
    \draw[Arrow] (fs) -- ++(-0.85, 0);
    \node[above] at ($(quantizer3) + (1, 0.675)$) {\tiny CLK};
    \node[rground] at (quantizer3.+) {};

    \draw (amp3.out) to[short, *-] (quantizer3.-);
    \draw (quantizer3.out) -- ++ (-0.5,0) to[R, l=$R_{\kappa_3}$] (vgnd3);

    \draw (vgnd3) -- ++(-0.5,0.5) to[R, l=$R_{\alpha_3}$] ++(0,1.75) node[inner sep=0] (alpha3) {};

    \draw (quantizer3.-) to[open, v^=$x_3(t)$] (quantizer3.+);
    \draw (quantizer3.out) to[open, v_=$s_3(t)$] ($(quantizer3.+) + (0,-0.785)$);

    \node[op amp] (ampN) at ($(amp3) + (5.875,0)$) {$A_N$};
    \node[rground] at (ampN.+) {};
    \node[inner sep=0] (vgndN) at ($(ampN.-) + (-0.5,-0.5)$) {};
    \draw (vgndN) to[short, *-] (ampN.-);
    \draw (vgndN) to[short, *-] ++(0,1.25) to[C, l=$C_N$] ++ (2.875,0) -| (ampN.out);

    \node[op amp, yscale=-1, rotate=180] (quantizerN) at ($(ampN) + (0, -1.5)$) {};
    \draw ($(quantizerN) + (0.5, -0.25)$) -- ++(-0.25,0) -- ++(0, 0.5) -- ++(-0.25, 0);
    \draw ($(quantizerN) + (0.1875, 0)$) -- ++(0.125,0);
    \node (fs) at ($(quantizerN) + (1.5, 0.675)$) {};
    \draw ($(fs) + (-0.125,-0.125)$) 
        -- ++(0.125, 0) 
        -- ++(0,0.25) 
        -- ++(0.125,0) 
        -- ++(0, -0.25) 
        -- ++(0.125, 0)
        -- ++(0, 0.25)
        -- ++(0.125, 0)
        -- ++ (0, -0.25)
        -- ++ (0.125, 0)
        -- ++ (0, 0.25)
        -- ++ (0.125, 0)
        -- ++ (0, -0.25)
        -- ++ (0.125,0 )
    ;
    \draw[{Stealth[length=0.875mm]}-{Stealth[length=0.875mm]}] ($(fs) + (-0.125, 0.25)$) -- node[above] {\tiny $T$} ++(0.25,0); 
    \draw[Arrow] (fs) -- ++(-0.85, 0);
    \node[above] at ($(quantizerN) + (1, 0.675)$) {\tiny CLK};
    \node[rground] at (quantizerN.+) {};

    \draw (ampN.out) to[short, *-] (quantizerN.-);
    \draw (quantizerN.out) -- ++ (-0.5,0) to[R, l=$R_{\kappa_N}$] (vgndN);

    \draw (quantizerN.-) to[open, v^=$x_N(t)$] (quantizerN.+);
    \draw (quantizerN.out) to[open, v_=$s_N(t)$] ($(quantizerN.+) + (0,-0.785)$);

    \draw (alpha1) -- ++(0, 0.5) -| ($(amp2.out) + (0.5,0.5)$) -- (amp2.out);
    \draw (alpha2) -- ++(0, 0.25) -| ($(amp3.out) + (0.5,0.5)$) -- (amp3.out);
    \draw (alpha3) -- ++(0, 0.5) -- ++(4,0) node[anchor=west] {\dots} ++(0.625,0) -| ($(ampN.out) + (0.5,0.5)$) -- (ampN.out);    

    \draw (amp1.out) to[R, l=$R_{\beta_2}$] (vgnd2);
    \draw (amp2.out) to[R, l=$R_{\beta_3}$] (vgnd3);
    \draw (amp3.out) -- ++(0.625,0) node[anchor=west] {\dots} ++(0.625,0) to[R, l=$R_{\beta_N}$] (vgndN);

    \draw ($(vgnd1) + (-2.25,-1.25)$) node[rground] {} to[V,l=$u(t)$] ++(0, 1.25) to[R,l=$R_{\beta_1}$] (vgnd1);
    \draw ($(vgnd1) + (-1.25,-2.875)$) node[rground] {} to[V,l=$s_0(t)$] ++(0, 1) to[R,l=$R_{\kappa_0}$] ++(0,1.25) -- (vgnd1);

\end{circuitikz}
        }
        \caption{\label{fig:leapfrog-structure}
        A single-ended circuit implementation of the $N$th order homogeneous \acf{LF} analog frontend with two input signals: $u(.)$, the unknown 
        signal to be converted, and $s_0(.)$, a known binary reference signal. 
        All signals $u(.), s_0(.), \dots, s_N(.), x_1(.), \dots, x_N(.)$ are represented as voltages.
        }
    \end{center}
\end{figure*}
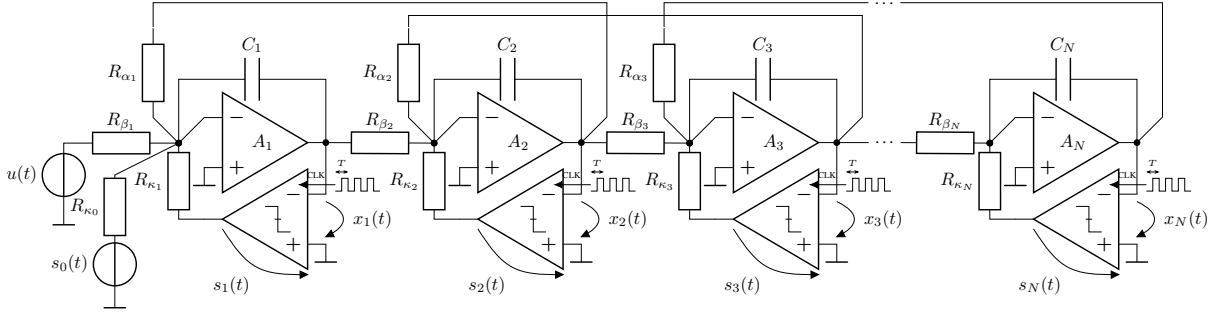
A single \ac{CBADC} analog frontend is considered for calibration; the \ac{CBADC} \ac{LF} analog frontend, introduced in \cite{M:20} and further investigated in \cite{FMWLY:2022}. 
A single-ended opamp-$RC$ based implementation of the $N$th order \ac{LF} structure is shown in \Fig{fig:leapfrog-structure}. Its basic functionality
can be thought of as amplifying an input signal $u(.)$ through a chain-of-integrators, each parametrized by a time constant $R_{\beta_\ell}C_\ell$,
while stabilizing the analog states $x_1(.),\dots,x_N(.)$ by local \ac{DC} signals $s_1(.),\dots,s_N(.)$. 
The analog frontend gets its name from the $R_{\alpha_\ell}$
feedback paths which also introduce complex poles in the transfer function from $u(.)$ to the final state $x_N(.)$
(assuming ($s_1(.)=0, \dots, s_N(.)=0$)). This transfer function is of fundamental importance for the performance of
a \ac{CBADC} and will be referred to by its impulse response $g_u(.)$.
The digital control is implemented as a clocked comparator, generating the binary discrete-time control signals $s_\ell[.]$ 
which are fed back to the system through a \ac{DAC} with an impulse response $\theta_\ell(.)$ as
\IEEEpubidadjcol
\begin{IEEEeqnarray}{rCl}
    s_\ell(t) & \eqdef &  \sum_k s_\ell[k] \theta_\ell(t - kT). \label{eq:continuos_to_discrete_time_control_signals}
\end{IEEEeqnarray}
The final state can be written as
\begin{IEEEeqnarray}{rCl}%
    x_N(t)  & = & (g_u \ast u)(t) + \sum_{\ell=0}^N (g_\ell \ast s_\ell)(t) \label{eq:magic_step}
\end{IEEEeqnarray}
where $g_\ell(.)$ denotes the impulse response of the transfer function from $s_\ell(.)$ to $x_N(.)$.

Finally, $s_0[k] \in \{\pm 1\}$ is a known binary discrete-time reference signal which will be used to calibrate the \ac{CBADC} frontend.
This signal is randomly generated to produce a wideband spectrum, and the ratio $R_{\kappa_0}/R_{\kappa_1}$ is a tradeoff parameter 
between the time needed for calibration and a reduced maximum input signal swing.

\section{The Digital Estimator}\label{sec:digital_estimator}
The output of the converter is an estimate $\hat u[k]$, of a filtered and sampled version of $u(.)$,  obtained by $N+1$ linear filters as
\begin{IEEEeqnarray}{rCl}
    \hat{u}[k]  & \eqdef & \sum_{\ell = 0}^N (h_\ell \ast s_\ell)[k] + \hat{u}_{0}  \label{eq:estimate}
\end{IEEEeqnarray}
where the reference filter $h_0$ is a design choice and $h_1, \dots, h_N$, and the scalar offset $\hat{u}_{0}$, 
will be subject to calibration.

As shown in \App{app:signal_decomposition}, 
$\hat{u}[.]$ can be decomposed into
\begin{IEEEeqnarray}{rCl}
    \hat{u}[k] & = & \left(h_u \ast u \right)(k T) + (\tilde{g}_u \ast x_N)(k T) + e_c[k]  \label{eq:estimate_decomposition}
\end{IEEEeqnarray}
where $h_u(.)$ is the impulse response of the desired \ac{STF}.
The second term in \Eq{eq:estimate_decomposition} is recognized as the nominal conversion error, 
where $\tilde{g}_u(.)$ is the impulse response of the \ac{NTF} \cite{M:20}.
Using the \ac{NTF} filter, the \ac{STF} filter can be witten as
\begin{IEEEeqnarray}{rCl}
    h_u(t) &=& - (\tilde{g}_u \ast g_u)(t) \label{eq:desired_filter}
\end{IEEEeqnarray}
with $g_u(.)$ as in \Eq{eq:magic_step}.
The last term in \Eq{eq:estimate_decomposition} is the calibration error
\begin{IEEEeqnarray}{rCl}%
    e_c[k] & \eqdef & \sum_{\ell = 1}^N \left(\left(h_\ell - \breve{h}_\ell\right) \ast s_\ell\right)[k] \labell{eq:calibration_error},%
\end{IEEEeqnarray}
where
\begin{IEEEeqnarray}{rCl}
    \breve{h}_\ell[k] & \eqdef &  (\tilde{g}_{u} \ast g_\ell \ast \theta_\ell)(kT) \label{eq:lmmse_filter},
\end{IEEEeqnarray}
are the minimizing filter coefficients to the cost function \Eq{eq:LMMSE} as 
will be further described in \Sec{sec:adaptive_filtering}.

The fundamental step in going from \Eq{eq:estimate} to \Eq{eq:estimate_decomposition} 
is connecting the desired continuous-time \ac{STF} filter and the discrete-time reference filter as
\begin{IEEEeqnarray}{rCl}%
    h_0[k] & \eqdef & \breve{h}_0[k] \propto - (h_u \ast \theta_0)(kT) \label{eq:reference_filter} %
\end{IEEEeqnarray}
where the proportionality relation is due to the fact that $g_u(.) \propto g_0(.)$.
The \ac{STF} filter is a digital design choice that is not subject to calibration. 
Moreover, the role of sampling, in connection to the \ac{STF} filter, is emphasized in \Eq{eq:estimate_decomposition} 
which opposes the conventional view that sampling is the work of the comparators in \Fig{fig:leapfrog-structure}. 
Despite the continuous-time nature of the \ac{STF} filter, 
it will be enforced by the digital estimator in the digital discrete-time domain, see \Eq{eq:reference_filter}.
Furthermore, 
as the \ac{NTF} filter is implicitly defined by the circuit implementation, i.e., $g_u(.)$ together with the choice of \ac{STF} filter, see \Eq{eq:desired_filter},
it is also not subject to calibration.

In summary,  \Eq{eq:estimate} and \Eq{eq:estimate_decomposition} reveals the fundamental idea of the control-bounded ADC concept in two equations:
a filtered and sampled version of the input signal, $\left(h_u \ast u \right)(k T)$, can be computed by a discrete-time convolution of
the control signals, \Eq{eq:estimate}, where the fundamental performance of such an estimate is dictated by the analog frontends 
ability to both amplify $u(.)$ %,i.e., $g_u(.)$, 
and bound the magnitude of the last state $x_N(.)$.
The calibration error $e_c[.]$ can be made arbitrarily small by conventional calibration techniques as will be shown next.

\subsection{Calibration}\label{sec:adaptive_filtering}
In the case of no input, i.e., $u(.) = 0$, 
the impulse responses $\breve{h}_\ell[.]$, from \Eq{eq:lmmse_filter}, are recognized, see \App{app:wiener_filter},
as the minimizing filter coefficients $h_1[.], \dots, h_N[.]$ with respect to the cost function
\begin{IEEEeqnarray}{rCl}%
    \EE{\left|(h_0 \ast s_0)[.] + \sum_{\ell = 1}^N (h_\ell \ast s_\ell)[.] + \hat{u}_{0}\right|^2}_{u(.)=0}  \label{eq:LMMSE}
\end{IEEEeqnarray}
where $s_0[.], \dots, s_N[.]$ are assumed to be zero-mean weakly stationary processes, 
jointly independent of the assumed zero-mean stationary process $(\tilde{g}_u \ast x_N)(kT)$, and $\EE{\cdot}$ denotes expectation.
This implies that the decomposition in \Eq{eq:estimate_decomposition} can be enforced, 
and the calibration error term made to vanish, by estimating $h_1, \dots, h_N$ from 
the discrete-time sequences $s_0[.], \dots, s_N[.]$ alone.
In particular, prior knowledge of any of the analog quantities $\tilde{g}_u(.)$, $g_u(.)$, $g_1(.), \dots, g_N(.)$, and $\theta_0(.),\dots,\theta_N(.)$ are immaterial
for such calibration. 

From here on, all filters $h_\ell[.]$ are chosen as FIR filter with equal filter length $K=512$.
Additionally, the reference filter $h_0[.]$ is determined by standard FIR filter design methods ($-3$ dB gain at bandwidth edge)
and its coefficient scaled by $-R_{\kappa_0}/R_{\beta_1}$ while all other $h_\ell[.]$ are initialized with all zero coefficients.
These choices of filter length, filter shape, and initialization are mainly motivated by simplicity and consistency of notation. 
Clearly, for a given application, better reference filter design choices could yield superior filtering performance at reduced complexity.

\subsubsection{Adaptive Filtering}\label{sec:LMS}
\begin{figure}
    \centering
    \resizebox{\columnwidth}{!}{%
    \begin{tikzpicture}[node distance=1.25cm]
    \node[Analog] (analog_frontend) {{analog frontend}};
    \node[Analog] (adaptive_filter) at ($(analog_frontend) + (4.75,0)$) {{adaptive filter}};
    \node[Analog] (desired_filter) at ($(analog_frontend) + (0, -1.5)$) {$h_0$};
    \node[AnalogSum] (asum) at ($(adaptive_filter) + (2, 0)$) {};
    \node[AnalogBranch] (abranch) at ($(asum) + (0.5, 0)$) {};
    \node (estimate) at ($(abranch) + (0.75,0) $) {$\hat{u}[k]$};
    \node[AnalogBranch] (abranch2) at ($(analog_frontend) + (0,-0.75)$) {};
    \node (s_0) at ($(abranch2) + (-2.25, 0)$) {$s_0[k]$};
    \node (u) at ($(s_0) + (-0.25, 0.75)$) {$u(t)=0$};

    \draw[Arrow] (analog_frontend) -- ++(2.475, 0) node[above] {$s_1[k], \dots, s_N[k]$} -- (adaptive_filter);
    \draw[Arrow] (s_0) -| (analog_frontend);
    \draw[Arrow] (abranch2) -- (desired_filter);
    \draw[Arrow] (abranch) -- (estimate);
    \draw (abranch) |- (asum);
    \draw[BackArrow] (asum) -- ++(-0.5, 0) -- (adaptive_filter);
    \draw[Arrow] (desired_filter) -| (asum);
    \draw[Arrow] (u) --  node[anchor=south] {} (analog_frontend);

    \draw[] (abranch) |- ($(adaptive_filter) + (-0.75, -1)$) -- ++(0.5, 0.625);
    \draw[Arrow] ($(adaptive_filter) + (0.25, 0.375)$) -- ++ (0.5, 0.625);

\end{tikzpicture}
    }
    \caption{\label{fig:adaptive_filter}Adaptive filtering setup for estimating $h_1, \dots, h_N$ using a 
    known reference signal $s_0[.]$ and a fixed reference filter $h_0$.
    }
\end{figure}
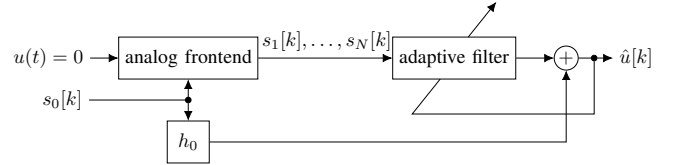
One way of estimating $h_1,\dots,h_N$ from \Eq{eq:LMMSE} is an adaptive filtering scheme as shown in \Fig{fig:adaptive_filter}. 
In this work, only frontend calibration, i.e., $u(.)=0$ during calibration, is considered. 
For the adaptive filter in \Fig{fig:adaptive_filter}, variations of the \ac{LMS} or \ac{RLS} algorithms \cite{Haykin:2002} can directly be applied.
An advantage of \ac{LMS} algorithm
is its low complexity as it only uses the gradient of \Eq{eq:LMMSE}, i.e., $2\EE{\vct{s}_\ell[.] \hat{u}[.]}$, in its update rule where
\begin{IEEEeqnarray}{rCl}
    \vct{s}_\ell[k] & \eqdef & \begin{pmatrix}s_\ell[k-K / 2], \dots, s_\ell[k + K / 2]\end{pmatrix}^\T \in \{\pm 1\}^{K}. \IEEEeqnarraynumspace
\end{IEEEeqnarray}

To decrease the number of calibration iterations, at the expense of additional computational and memory complexity,
the \ac{RLS} algorithm \cite{Haykin:2002,LDHKPK:2007} is a viable alternative.
For simplicity, only the \ac{RLS} algorithm was considered in this work.
The \ac{RLS} algorithm minimizes the cost function
\begin{IEEEeqnarray}{rCl}
    \argmin_{\vct{h}} \, \sum_{k=0}^{K_i} \lambda^{K_i-k} \left|\hat{u}[k]\right|^2 + \lambda^K_i \delta \|\vct{h}^0\|_2^2, \label{eq:RLS_cost_function}
\end{IEEEeqnarray}
for $0 < \lambda \leq 1$, and $\delta \geq 0$, which converge to \Eq{eq:LMMSE} for $\lambda < 1$ and the number of calibration iterations $K_i \to \infty$. % Notice that for $\lambda = 1$ and $\delta = 0$ 
The algorithm proceeds recursively as
\begin{IEEEeqnarray}{rCl}
    \vct{\alpha}_k & = & \vct{V}^{k-1}\vct{s}[k] \\
    \vct{g}_k & = & \vct{\alpha}_k \left(\lambda + \vct{s}[k]^\T \vct{\alpha}_k \right)^{-1} \\
    \vct{V}^k & = & \frac{1}{\lambda}\left(\vct{V}^{k-1} - \vct{g}_k \vct{\alpha}_k^\T \right) \\
    \vct{h}^{k} & = & \vct{h}^{k-1} - \hat{u}[k]\vct{g}_k
\end{IEEEeqnarray}
where
\begin{IEEEeqnarray}{rCl}%
    \vct{s}[.] & \eqdef & \begin{pmatrix}\vct{s}_1[.]^\T, \dots, \vct{s}_N[.]^\T, 1\end{pmatrix}^\T\in \{\pm 1\}^{K_{\Sigma}}, \label{eq:s_vector} \\
    \vct{h}[.] & \eqdef & \begin{pmatrix}h_1[.]^\T, \dots, h_N[.]^\T, \hat{u}_{0}\end{pmatrix}^\T \in \R^{K_{\Sigma}}, \label{eq:vector_h}
\end{IEEEeqnarray}
$K_{\Sigma} = N K + 1$, $\vct{\alpha}_k\in \R^{K_{\Sigma}}$, $\vct{g}_k\in \R^{K_{\Sigma}}$, and $\vct{V}^{k} \in \R^{K_{\Sigma} \times K_{\Sigma}}$ is a symmetric matrix.
The regularization in \Eq{eq:RLS_cost_function} is enforced by initializing the matrix $\vct{V}^0 = \delta^{-1} \vct{I}_{K_{\Sigma}}$ where $\vct{I}_{K_{\Sigma}}$ is the $K_{\Sigma}$-by-$K_{\Sigma}$ identity matrix.
Throughout this paper, $\lambda = 1.0 - 10^{-12}$ and $\delta = 0.01$ for all simulations.

\section{Simulation Results}\label{sec:simulations}
The circuit from \Fig{fig:leapfrog-structure} is
parameterized, in accordance with \cite{FMWLY:2023}, for a nominal performance of 76.5~dB \ac{SNR} at 10~MHz signal bandwidth and a system order of $N=6$. 
Specifically, the time constants are chosen as follows: $R_{\alpha_{\ell}}C_{\ell} = 98.63$~ns, $R_{\beta_{\ell}}C_\ell = 10.30$~ns, 
$R_{\kappa_\ell}C_\ell=4 R_{\beta_{\ell}}C_\ell$, for every $\ell$ with the exceptions of  
$R_{\beta_1}C_1 = 2 R_{\beta_{\ell \neq 1}} C_{\ell \neq 1}$. 
This results in a maximum input signal swing that is twice that of the maximum state swing and half that of the maximum control
signal swing. 
The comparators in \Fig{fig:leapfrog-structure} are clocked
at $f_s = 194.7$~MHz resulting in an \ac{OSR} of approximately $10$.

\subsection{Behavioral Simulation}\label{sec:behavioral-simulation}
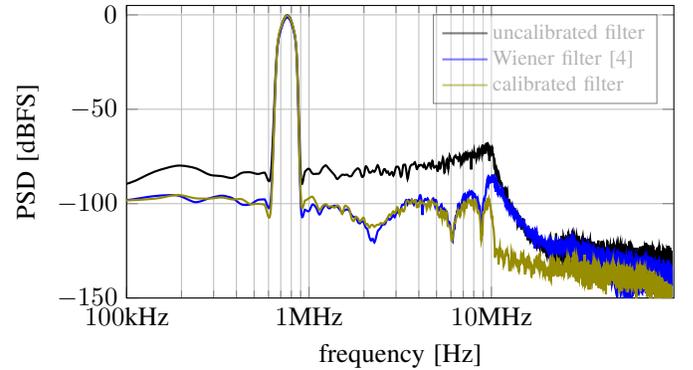
\begin{figure}
    \centering
    \begin{tikzpicture}
        \pgfplotsset{
            grid=both, 
        }
        \pgfplotsset{ylabel near ticks, xlabel near ticks}
        \pgfplotsset{ 
                legend cell align={left},
                legend style={
                    at={(0.985,0.98)},
                    anchor=north east,
                    fill opacity=0.3,
                    draw opacity=0.5,
                    text opacity=1,
                    nodes={scale=0.8, transform shape}
                }
        }
        \begin{semilogxaxis}[
            xlabel={frequency [Hz]},
            ylabel={PSD [dBFS]},
            ymajorgrids,
            xminorgrids,            
            xmin=1e5,
            xmax=1e8,
            xtick = {1e5,1e6,1e7},
            xticklabels = {$100$kHz,$1$MHz,$10$MHz},
            width=\columnwidth,
            height=\columnwidth/1.618,
            legend pos = north east,
            ymax=5,
            ymin=-150,
            ]
            \addplot[color=black, thick, smooth] table[col sep=comma, x=f_unc_luna_batch_3_rls_K_9, y expr={\thisrow{unc_luna_batch_3_rls_K_9} + 65}]{./figures/psd_component_variation.csv};
            \addplot[color=blue, thick, smooth] table[col sep=comma, x=f_ref_luna_batch_3_rls_K_9, y expr={\thisrow{ref_luna_batch_3_rls_K_9} + 65}]{./figures/psd_component_variation.csv};
            \addplot[color=olive, thick, smooth] table[col sep=comma, x=f_cal_luna_batch_3_rls_K_9, y expr={\thisrow{cal_luna_batch_3_rls_K_9} + 65}]{./figures/psd_component_variation.csv};

            \addlegendentry{uncalibrated filter}
            \addlegendentry{Wiener filter \cite{MWL:21}}
            \addlegendentry{calibrated filter}
        \end{semilogxaxis}
    \end{tikzpicture}
    \caption{\label{fig:component_variation_psd}A representative \ac{PSD} of \Eq{eq:estimate} 
    from one of the realizations in \Sec{sec:behavioral-simulation}.}
\end{figure}

A component variation scenario is considered
where all time constants 
$(R_{\kappa_0} C_1)$, $(R_{\alpha_\ell} C_\ell)$, $(R_{\beta_\ell} C_\ell)$, and $(R_{\kappa_\ell}C_\ell)$
are subject to variations from their nominal values. 
Ideal amplifiers, comparators, a non-return to zero \ac{DAC}, and passive components were modeled in Verilog-A, and using the Spectre simulator, 
128 Monte Carlo simulations were conducted were each of the mentioned time constants are drawn 
independently and uniformly at random within $\pm 10\%$ of their nominal values.
Every sample circuit is simulated with (testing) and without (training) a test sinusoidal input signal and 
the calibration is then conducted on the latter, and validated on the former, dataset. 
The test signal has a frequency of $f_s / 2^8 \approx 760$~kHz and an amplitude of $-1$~\ac{dBFS} to avoid potential 
instability due to the component variations. We use $R_{\kappa_1} / R_{\kappa_0} = 0.1$ and the Wiener filter
from \cite{MWL:21} (with $\eta^2$ chosen as (22) in \cite{MWL:21}), with the true time constants, 
as a reference for the calibration results.
\begin{figure}
    \centering
    \begin{tikzpicture}
        \pgfplotsset{
            grid=both, 
        }
        \pgfplotsset{ylabel near ticks, xlabel near ticks}
        \pgfplotsset{ 
            legend cell align={left},
            legend style={
                at={(0.985,0.98)},
                anchor=north east,
                fill opacity=0.9,
                draw opacity=1,
                text opacity=1,
                nodes={scale=0.8, transform shape}
        }
        }
        \begin{semilogxaxis}[
            xlabel={number of calibration iterations $K_i$},
            ylabel={Testing SNR [dB]},
            ymajorgrids,
            yminorgrids,
            xminorgrids,            
            xmin=2048,
            xmax=16384,
            ymin=20,
            ymax=85,
            legend pos = south east,
            width=\columnwidth,
            height=\columnwidth/1.618,
            mark size=0.75,
            xtick = {1024, 2048, 4096, 8192, 16384, 32768, 65536},
            xticklabels = {$2^{10}$, $2^{11}$, $2^{12}$, $2^{13}$, $2^{14}$, $2^{15}$, $2^{16}$},
            ytick = {20, 30, 40, 50, 60, 70, 80, 90},
            ]
            \addplot[dashed, thick] table[col sep=comma, x=i, y=avg_uncal] {./figures/ref_uncal_firwin2_K512.csv};
            \addplot[color=blue, dash dot, thick] table[col sep=comma, x=i, y=avg_ref] {./figures/ref_uncal_firwin2_K512.csv};
            \addplot[color=red, thick] table[col sep=comma, x=i, y=avg] {./figures/testing_error_component_variation.csv};
            \addlegendentry{average uncalibrated filter}
            \addlegendentry{average Wiener filter \cite{MWL:21}}
            \addlegendentry{average calibrated filter}

            \addplot[name path=uu,color=gray, opacity=0] table[col sep=comma, x=i, y=max_uncal] {./figures/ref_uncal_firwin2_K512.csv};
            \addplot[name path=ul,color=gray, opacity=0] table[col sep=comma, x=i, y=min_uncal] {./figures/ref_uncal_firwin2_K512.csv};
            \addplot[gray,opacity=0.45] fill between[of=uu and ul];

            \addplot[color=blue, name path=ru,color=gray, opacity=0] table[col sep=comma, x=i, y=max_ref] {./figures/ref_uncal_firwin2_K512.csv};
            \addplot[color=blue, name path=rl,color=gray, opacity=0] table[col sep=comma, x=i, y=min_ref] {./figures/ref_uncal_firwin2_K512.csv};
            \addplot[color=blue, opacity=0.3] fill between[of=ru and rl];

            \addplot[
                color=red,
                opacity=0,
                name path=u_rsl, 
            ] table[col sep=comma, x=i, y=max] {./figures/testing_error_component_variation.csv};
            \addplot[
                color=red,
                opacity=0,
                name path=l_rsl,
            ] table[col sep=comma, x=i, y=min] {./figures/testing_error_component_variation.csv};
            \addplot[red, opacity=0.25] fill between[of=u_rsl and l_rsl];

        \end{semilogxaxis}
    \end{tikzpicture}
    \caption{\label{fig:convergence}The testing error performance for the uncalibrated and reference case
    as well as the \ac{RLS} algorithm convergence behaviour as a function of the number of calibration iterations. 
    The colored areas, associated with each case, represent the range of testing performance over all 128 sample circuits.}
\end{figure}
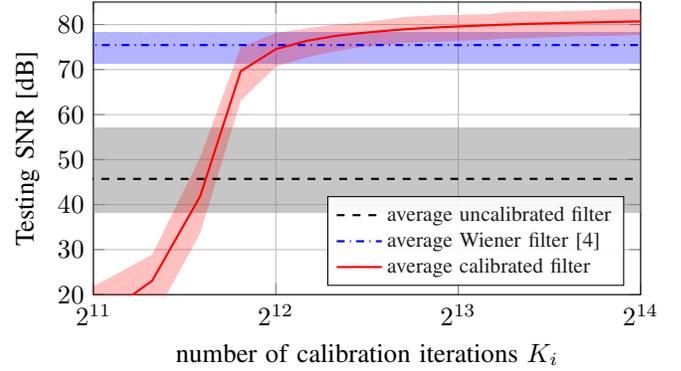
\Fig{fig:convergence} shows the average testing performance for the uncalibrated Wiener filter (with nominal filter coefficients),
the reference Wiener filter (using true filter coefficients), and the \ac{RLS} algorithm's testing performance as a function
of the number of calibration iterations. 
As stated in \cite{FMWLY:2022,FMWLY:2023}, without calibration, the digital estimator suffers significant performance degradation due to mismatch between
the digital estimator and the, assumed nominal, analog frontend parametrization. Clearly, the mismatch can be managed as the \ac{RLS} algorithm
reaches the reference filter performance after $\approx 2^{12}$ iterations.
The leapfrog structure is known for its resilience towards component variations,
which is confirmed by the (average, maximum, minimum) SNR performance of $(80.70,83.51,77.68)$ and $(75.45,78.38,71.34)$ dB
for the calibrated filter and reference Wiener filter respectively.
\Fig{fig:component_variation_psd} shows the \ac{PSD}, of \Eq{eq:estimate}, from a representative sample of the testing dataset.
The discrepancy around the bandwidth frequencies, between the Wiener reference and the calibrated filter, is due to the
$h_0[.]$ being designed as standard FIR filter and not the Wiener filter from \cite{MWL:21}. 
As the \ac{SNR} is determined directly from the \ac{PSD}, 
this also explains why the calibrated filter outperforms the reference filter in \Fig{fig:convergence}.

\subsection{Transistor Level Simulation} \label{sec:example}
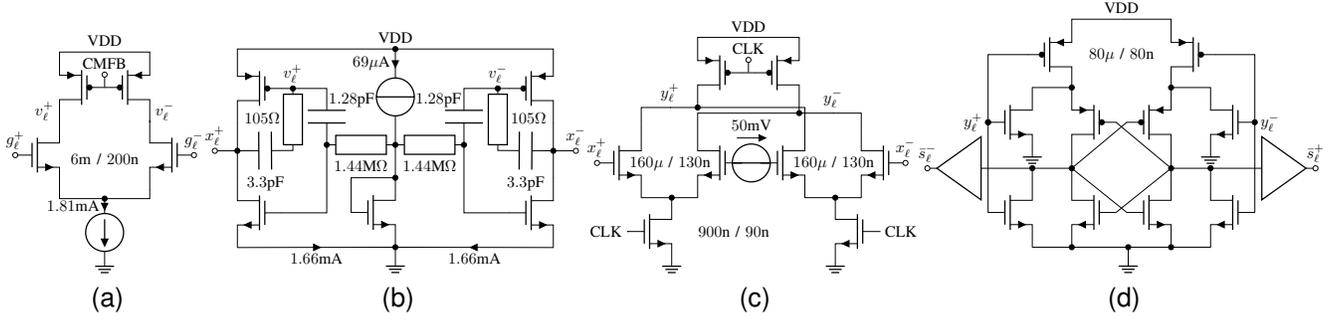
\begin{figure*}
    \centering
    \subfloat[][]{
        \label{subfig:transistor_level_diff_pair}
        \scalebox{0.6}{
            \begin{circuitikz}[european voltages]
    \ctikzset{tripoles/mos style=arrows}
    \draw(0,0) coordinate (vcc) node[above] {$\text{VDD}$};
    \node[pmos, xscale=-1] (q1p) at (-1,-0.785) {};
    \node[pmos] (q1n) at (1,-0.785) {};
    \draw (vcc) to[short,-] ++(0,0) -| (q1p.S);
    \draw (vcc) -| (q1n.S);
    \draw(q1p.G) -- (q1n.G);
    \node (CMFB) at ($(vcc) + (0,-0.25)$) {$\text{CMFB}$};
    \draw ($(CMFB)+(0,-0.25)$) to[short,o-] ++(0,-0.275);
    \draw (q1p.D)  ++(-0, 0.25) node[left] {$v^+_{\ell}$};
    \draw (q1n.D)  ++(0, 0.25) node[right] {$v^-_{\ell}$};

    \node[nmos] (q2p) at ($(q1p) + (0,-1.5)$) {};
    \node[nmos, xscale=-1] (q2n) at ($(q1n) + (0,-1.5)$) {};
    \draw(q1p.D) -- (q2p.D);
    \draw(q1n.D) -- (q1n.D);
    \node[above] (v_in_p) at (q2p.G) {$g^+_\ell$};
    \node[above] (v_in_n) at (q2n.G) {$g^-_\ell$};
    \draw (q2p.G) to[short, o-] (q2p.G);
    \draw (q2n.G) to[short, o-] (q2n.G);

    \node at ($(vcc) + (0,-2.375)$) {$6$m / $200$n};
    
    \draw ($(vcc) + (0,-3.25)$) coordinate (ibv) to[isource, i>_=$1.81$mA, *-] ++(0,-1.5) node[tlground] {};
    \draw (q2n.S) |- (ibv) -| (q2p.S);
\end{circuitikz}
        }
    }
    \hspace{-0.7cm}
    \subfloat[][]{
        \label{subfig:transistor_level_buffer}
        \scalebox{0.6}{
            \begin{circuitikz}[european]
    \ctikzset{tripoles/mos style=arrows}
    \draw(0,0) coordinate (vcc) node[above] {$\text{VDD}$};
    \node[pmos, xscale=-1] (q1p) at (-3.5,-0.875) {};
    \node[pmos] (q1n) at (3.5,-0.875) {};
    \draw (vcc) to[short, *-] ++(0,0) -| (q1p.S);
    \draw (vcc) -| (q1n.S);
    \draw ($(q1p.G)+(0.25,0)$) to[short, *-] ++(0,0) node[above] (v_in_p) {$v^+_{\ell}$};
    \draw ($(q1n.G)+(-0.25,0)$) to[short, *-] ++(0,0) node[above] (v_in_n) {$v^-_{\ell}$};

    \coordinate (voutp) at ($(q1p.D) + (0,-0.6125)$);
    \draw (voutp) to[short, *-o] ++(-0.5, 0) node[above] {$x^+_\ell$};

    \coordinate (voutn) at ($(q1n.D) + (0,-0.6125)$);
    \draw (voutn) to[short, *-o] ++(0.5, 0) node[above] {$x^-_\ell$};

    \draw (vcc) to[isource, i>_=$69\mu$A] ++(0,-2.125) coordinate (ib){};
    \node[nmos] (q3) at ($(ib) + (0,-1.5)$) {};
    \draw (q3.G) |- (q3.D) to[short, *-*] (ib);
    
    \node[nmos, xscale=-1] (q2p) at ($(voutp) + (0,-1.375)$) {};
    \node[nmos] (q2n) at ($(voutn) + (0,-1.375)$) {};
    \draw(q1p.D) -- (q2p.D);
    \draw(q1n.D) -- (q2n.D);

    \draw (q1p.G) to[short] ++ (0.25,0) coordinate (vptemp1) to[R,l_=$105\Omega$] ++ (0,-1.375) to[C,l=$3.3$pF] (voutp);
    \draw (vptemp1) to[short] ++(0.75,0) to[C] ++(0,-1.25) coordinate (vptemp2) to[R,l_=$1.44\text{M}\Omega$, *-] (ib);
    \draw (vptemp2) |- (q2p.G);
    \node[right] at ($(vptemp2)+(-0.075,1.0)$) {$1.28$pF};

    \draw (q1n.G) to[short] ++ (-0.25,0) coordinate (vntemp1) to[R,l=$105\Omega$] ++ (0,-1.375) to[C,l_=$3.3$pF] (voutn);
    \draw (vntemp1) to[short] ++(-0.75,0) to[C] ++(0,-1.25) coordinate (vntemp2) to[R,l=$1.44\text{M}\Omega$, *-] (ib);
    \draw (vntemp2) |- (q2n.G);
    \node[left] at ($(vntemp2)+(0.05,1.0)$) {$1.28$pF};

    \draw (q3.S) node[ground] {} to[short, *-] ++(0,0);
    \draw (q2n.S) to[short,i=$1.66$mA] (q3.S);
    \draw (q2p.S) to[short,i_=$1.66$mA] (q3.S);

\end{circuitikz}
        }
    }
    \hspace{-0.7cm}
    \subfloat[][]{
        \label{subfig:miyahara_comparator}
        \scalebox{0.6}{
            \begin{circuitikz}[european]
    \ctikzset{tripoles/mos style=arrows}
    \draw(0,0) coordinate (vcc) node[above] {$\text{VDD}$};
    \node[pmos, xscale=-1] (q1p) at (-1.125,-0.785) {};
    \node[pmos] (q1n) at (1.125,-0.785) {};
    \draw (vcc) to[short,-] ++(0,0) -| (q1p.S);
    \draw (vcc) -| (q1n.S);
    \draw (q1n.G) to[short] (q1p.G);
    \node at ($(vcc) + (0,-0.25)$) {$\text{CLK}$};
    \draw ($(vcc) + (0,-0.5)$) to[short,o-] ++(0,-0.275);
    \node[nmos] (q2p) at ($(q1p) + (-1.25,-2)$) {};
    \node[nmos, xscale=-1] (q2n) at ($(q1p) + (0,-2)$) {};
    
    \node[nmos] (q3p) at ($(q1n) + (0.125,-2)$) {};
    \node[nmos, xscale=-1] (q3n) at ($(q1n) + (1.375,-2)$) {};
    
    \draw (q2n.G) to[vsource, v=$50$mV] (q3p.G);

    \draw(q1p.D) to[short, -*] ++(0,0) coordinate (temp1) -| (q2p.D);
    \draw(temp1) -| (q3p.D);

    \draw(q1n.D) to[short, -*] ++(0,-0.125) coordinate (temp2) -| (q2n.D);
    \draw(temp2) -| (q3n.D);

    \draw ($(temp1) + (-0.6125,0)$) to[short,-] ++(0,0) node[above] {$y^+_\ell$};
    \draw ($(temp2) + (0.75,0)$) to[short,-] ++(0,0) node[above] {$y^-_\ell$};

    \draw (q2p.S) to[short,-*] ++(0.675,0) coordinate (temp3) to[short] (q2n.S);
    \node[nmos] (q4) at ($(temp3) + (0,-0.75)$) {};
    \draw (temp3) to[short] (q4.D);
    \draw (q4.S) node[tlground] {};
    
    \draw (q3p.S) to[short,-*] ++(0.675,0) coordinate (temp4) to[short] (q3n.S);
    \node[nmos, xscale=-1] (q5) at ($(temp4) + (0,-0.75)$) {};
    \node[] at ($(q5) + (-2.25,0)$) {$900$n / $90$n};
    \draw (temp4) to[short] (q5.D);
    \draw (q5.S) node[tlground] {};

    \draw (q2p.G) to[short,-o] ++(0, 0) node[above] {$x^+_\ell$};   
    \draw (q3n.G) to[short,-o] ++(0, 0) node[above] {$x^-_\ell$};

    \draw (q4.G) to[short,-] ++(0, 0) node[left] {$\text{CLK}$};
    \draw (q5.G) to[short,-] ++(0, 0) node[right] {$\text{CLK}$};
    
    \node[] at ($(q2n) + (-0.6125,0)$) {$160\mu$ / $130$n};
    \node[] at ($(q3n) + (-0.6125,0)$) {$160\mu$ / $130$n};
    
\end{circuitikz}
        }
    }
    \hspace{-0.7cm}
    \subfloat[][]{
        \label{subfig:miyahara_comparator_2}
        \scalebox{0.6}{
            \begin{circuitikz}[european voltages]
    \ctikzset{tripoles/mos style=arrows}
    \draw(0,0) coordinate (vcc) node[above] {$\text{VDD}$};
    \node[pmos] (q1p) at (-1.1,-0.785) {};
    \node[pmos,xscale=-1] (q1n) at (1.1,-0.785) {};
    \draw (vcc) to[short,-] ++(0,0) -| (q1p.S);
    \draw (vcc) -| (q1n.S);
    
    \draw (q1n.G) to[short] ++(0,0) node[right] {};
    \draw (q1p.G) to[short] ++(0,0) node[left] {};

    \node[] at ($(vcc) + (0,-0.785)$) {$80\mu$ / $80$n};

    \node[nmos] (q211) at ($(q1p) + (-.875,-1.5)$) {};
    \node[pmos,xscale=-1] (q212) at ($(q1p) + (0,-1.5)$) {};
    \draw (q211.D) to[short,-*] (q212.S);
    \draw (q211.G) to[short,-o] ++(0,0) node[left] {$y^+_\ell$};
    \draw (q211.S)  node[tlground] {};

    \node[nmos] (q221) at ($(q211)+(0,-2)$) {};
    \draw (q221.D) to[short,-*] ++(0,0.2125);

    \node[nmos,xscale=-1] (q222) at ($(q212)+(0,-2)$) {};
    \draw (q222.D) to[short] (q212.D);

    \node[nmos, xscale=-1] (q311) at ($(q1n) + (.875,-1.5)$) {};
    \node[pmos] (q312) at ($(q1n) + (0,-1.5)$) {};
    \draw (q311.D) to[short,-*] (q312.S);
    \draw (q311.G) to[short] ++(0,0) node[right] {$y^-_\ell$};
    \draw (q311.S)  node[tlground] {};

    \node[nmos,xscale=-1] (q321) at ($(q311)+(0,-2)$) {};
    \draw (q321.D) to[short,-*] ++(0,0.2125);

    \node[nmos] (q322) at ($(q312)+(0,-2)$) {};
    \draw (q322.D) to[short] (q312.D);

    \coordinate (sp) at ($(q212.D) + (0,-0.25)$);
    \coordinate (sn) at ($(q312.D) + (0,-0.25)$);
    
    \node[buffer, xscale=-1] (buf1) at ($(sp) + (-2.5,0)$) {};
    \draw (buf1.in) to[short, -*] (sp);

    \node[buffer, xscale=1] (buf2) at ($(sn) + (2.5,0)$) {};
    \draw (buf2.in) to[short, -*] (sn);

    \draw (q212.G) to[short] (sn);
    \draw (q222.G) to[short] (sn);
    \draw (q312.G) to[short] (sp);
    \draw (q322.G) to[short] (sp);

    \draw (buf1.out) to[short,-o] ++(0,0) node[above] {$\bar{s}_\ell^-$};
    \draw (buf2.out) to[short,-o] ++(0,0) node[above] {$\bar{s}_\ell^+$};

    \draw (q221.S) to[short] (q222.S) to[short, *-*] (q322.S) to[short] (q321.S);
    \draw ($(q222.S) + (1.25,0)$) to[short, -*] ++(0,0) node[ground] {};

    \draw (q221.G) to[short, -*] (q211.G) |- (q1p.G); 
    \draw (q321.G) to[short, -*] (q311.G) |- (q1n.G); 
    
\end{circuitikz}
        }
    }
    \caption{\label{fig:transistor_level}Transistor-level implementation: (a) the input stage of the first op-amp from \Fig{fig:leapfrog-structure} where $(g^+_{\ell}, g^-_{\ell})$
    are the corresponding positive and negative virtual grounds, (b) the output buffer stage of the first op-amp, (c) the decision stage, and (d) the output stage of the Miyahara comparator. Note that an additional latch circuit, not included in (d), connects $(\bar{s}_\ell^+, \bar{s}_\ell^-)$ to the control outputs $(s_\ell^+, s_\ell^-)$.
    }
\end{figure*}

To further validate the calibration algorithm, a differential transistor-level implementation of \Fig{fig:leapfrog-structure} was designed and simulated with Spectre in a 65-nm CMOS technology using the parametrization of \Sec{sec:simulations}. The design was purposefully over-dimensioned in terms of linearity such that thermal noise levels limit the performance. 
The specifics are as follows: 1.2~V power supply, $C_1=11.7$~pF, $C_2=C_3=2.925$~pF, and $C_4=C_5=C_6=975$~fF, a differential input signal of $1.2$~Vpp and $R_{\kappa_1} / R_{\kappa_0} = 0.03$.
The amplifiers are class AB two-stage RC-compensated amplifiers as in \cite{RACGLM:2006}.
\Fig{fig:transistor_level} (a) and (b) shows the internals of the first amplifier in \Fig{fig:leapfrog-structure}.
The first stage amplifier has an open-loop simulated input-referred noise level of $24.9~\mu$Vrms over the $10$~MHz signal bandwidth. 
Furthermore, the first stage amplifier, when configured as an inverting op-amp with a capacitive feedback $C_1$ and a resistive input $R_{\beta_1}$, measures a 83.8~dB \ac{SDR} for a full-scale input sinusoidal signal at 0.76~MHz.
The amplifiers are downsized to somewhat match the diminishing thermal noise and linearity requirements. 
The approximate scaling can be seen from the resulting
simulated power consumption as $P_{A_1}=8.5$~mW, $P_{A_2} = P_{A_3}=2.3$~mW, and $P_{A_4}=\dots=P_{A_6}=0.9$~mW.
The comparators, see \Fig{fig:transistor_level} (c) and (d), are based on \cite{MAPM:2008} and have an input referred noise of $130$~$\mu$Vrms which was achieved by extending the decision time to a quarter of the control period. 
Stability issues, due to long decision time, were addressed by using ternary comparators with decision thresholds at $\pm50$~mV. 
The ternary comparators result in a reduced maximum state swing, by approximately $30\%$, which translates into $\approx3$~dB \ac{SNR} gain, see \Eq{eq:estimate_decomposition}.
Some characteristic simulated waveforms from the first and second stage of the designed \ac{CBADC} can be seen in \Fig{fig:waveforms}.

In \Fig{fig:psd}, both the calibrated and uncalibrated PSDs of the estimate from \Eq{eq:estimate} are shown for a FS input signal at $f_s/2^8 \approx 0.76$~MHz.
A \ac{SNDR} of $80.4$ (calibrated) and $58.3$ (uncalibrated)~dB % and a \ac{SFDR} of $\approx97$~dB 
is estimated directly from the \ac{PSD}. 
The transistor level implementation has a simulated power consumption of $18.43$~mW.

\begin{figure}
    \centering
    \begin{tikzpicture}
    \pgfplotsset{
        grid=both, 
    }
    \pgfplotsset{ylabel near ticks, xlabel near ticks}
    \pgfplotsset{ 
        legend cell align={left},
        legend style={
            at={(0.985,0.98)},
            anchor=north east,
            fill opacity=1,
            draw opacity=1,
            text opacity=1,
            nodes={scale=0.8, transform shape}
    }
    }
    \begin{groupplot}[
        group style={
          group size=1 by 2,
          vertical sep=5pt,
          group name=G},
        xmin=50e-9,
        xmax=130e-9,
        xtick={60e-9, 80e-9, 100e-9, 120e-9},
        xticklabels={$60$ns, $80$ns, $100$ns, $120$ns},
        scaled ticks=false,
        name=state,
        ytick=\empty,
        width=0.8\columnwidth,
        axis lines=left,
        no markers,
        legend pos=outer north east,
        ]

        \nextgroupplot[
                xticklabels={},
                ymin=-1.4,
                ymax=1.4,
                xmin=50e-9,
                xmax=130e-9,
                xtick={60e-9, 80e-9, 100e-9, 120e-9},
                ytick={-1,0,1},
                height=3cm,
            ]
        \addplot[thick] table[col sep=comma, x=x, y=y] {./figures/signals/clk_short.csv};
        \addplot[cyan, thick] table[col sep=comma, x=x, y=y] {./figures/signals/s_0_short.csv};
        \addplot[orange, thick] table[col sep=comma, x=x, y=y] {./figures/signals/s_1_short.csv};
        \addplot[blue, thick] table[col sep=comma, x=x, y=y] {./figures/signals/s_2_short.csv};
        \addplot[red] table[col sep=comma, x=x, y=y] {./figures/signals/u_short.csv};
        \addlegendentry{clk}
        \addlegendentry{$s_0(t)$}
        \addlegendentry{$s_1(t)$}
        \addlegendentry{$s_2(t)$}
        \addlegendentry{$u(t)$}

        \nextgroupplot[
            ymin=-0.1,
            ymax=0.125,
            ytick={-0.2,-0.1,0,0.1,0.25},
            height=3.5cm,
        ]
        \addplot[red] table[col sep=comma, x=x, y=y] {./figures/signals/x_1_short.csv};
        \addplot[olive] table[col sep=comma, x=x, y=y] {./figures/signals/x_2_short.csv};
        \addlegendentry{$x_1(t)$}
        \addlegendentry{$x_2(t)$}

    \end{groupplot}
    \node[rotate=90,xshift=-0.25cm, yshift=1cm] {Differential Voltage};
\end{tikzpicture}
    \caption{\label{fig:waveforms}Simulated waveforms of the differential input, control signals, and first and second state variables of the simulated transistor-level implementation.}
\end{figure}
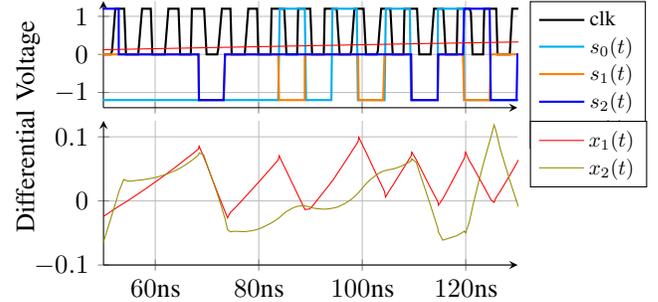
\begin{figure}
    \centering
    \begin{tikzpicture}
        \pgfplotsset{
            grid=both, 
        }
        \pgfplotsset{ylabel near ticks, xlabel near ticks}
        \pgfplotsset{ 
                legend cell align={left},
                legend style={
                    at={(0.985,0.98)},
                    anchor=north east,
                    fill opacity=0.3,
                    draw opacity=0.5,
                    text opacity=1,
                    nodes={scale=0.8, transform shape}
                }
        }
        \begin{semilogxaxis}[
            xlabel={frequency [Hz]},
            ylabel={PSD [dBFS]},
            ymajorgrids,
            xminorgrids,            
            xmin=1e5,
            xmax=1e8,
            ymin=-150,
            ymax=5,
            xtick = {1e5,1e6,1e7},
            xticklabels = {$100$kHz,$1$MHz,$10$MHz},
            width=\columnwidth,
            height=\columnwidth/1.618,
            legend pos=north east
            ]
            \addplot[color=black,thick, smooth] table[col sep=comma, x=f_unc_700k_rls_K_9_v1, y expr={\thisrow{unc_700k_rls_K_9_v1} + 65}] {./figures/psd_circuit.csv};
            \addplot[color=blue,thick, smooth] table[col sep=comma, x=f_cal_700k_rls_K_9_v1, y expr={\thisrow{cal_700k_rls_K_9_v1} + 65}] {./figures/psd_circuit.csv};

            \addlegendentry{uncalibrated filter} %(60.8dB SNDR)
            \addlegendentry{calibrated filter} %(80.8dB SNDR)

            \coordinate (third_harmonic_top) at (axis cs:4.56328125e6,-63) {};
            \coordinate (third_harmonic_bottom) at (axis cs:4.56328125e6,-160) {};
        \end{semilogxaxis}
    \end{tikzpicture}
    \caption{\label{fig:psd}Uncalibrated (58.3~dB SNDR) and calibrated (80.4~dB SNDR) PSDs of \Eq{eq:estimate} 
    from the transistor level simulation in \Sec{sec:example}. 
    }
\end{figure}
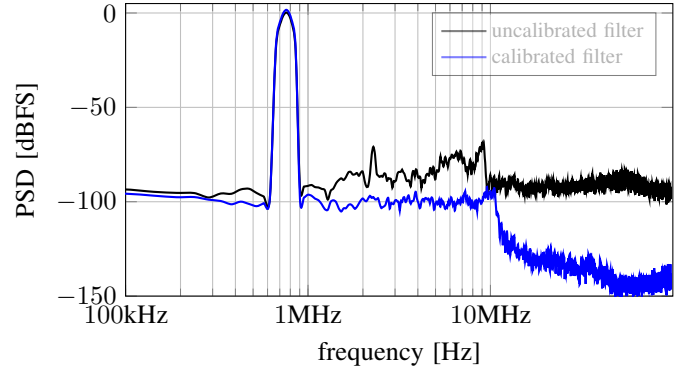

\section{Conclusions}\label{sec:conlusions}
Both theoretical and practical aspects of calibrating \acp{CBADC} were addressed. The proposed calibration
algorithm was validated using both behavioral and transistor level simulations, confirming
recovery of essentially nominal performance. 
The introduction of calibration to the \ac{CBADC} concept splits the \ac{CBADC} design task into two disjoint parts:
an analog frontend where the fundamental performance follows from analog design parameters and
a digital calibration step that is agnostic to the analog implementation.

\appendix \label{app:app}

\subsection{The Signal Estimate Decomposition}\label{app:signal_decomposition}

The discrete-time convolution can be written as 
\begin{IEEEeqnarray}{rCl}
    \IEEEeqnarraymulticol{3}{l}{
        (\breve{h}_\ell \ast s_\ell)[k]}  \nonumber\\* \quad
               \eqdef && \sum_{k_1} s_\ell[k_1] \breve{h}_\ell[k - k_1] \label{eq:discrete_continuous_first}  \\
                 = && \sum_{k_1} s_\ell[k_1] (\tilde{g}_u \ast g_\ell \ast \theta_\ell)((k - k_1) T)  \label{eq:discrete_continuous_second}  \\
                 = && \int (\tilde{g}_u \ast g_\ell)(\tau)  \sum_{k_1} s_\ell[k_1] \theta_\ell(kT - \tau - k_1 T ) \dd \tau \label{eq:sum_integral_change} \\
                = && \int (\tilde{g}_u \ast g_\ell)(\tau) s_\ell(kT - \tau) \dd \tau \label{eq:discrete_continuous_second_to_last} \\
                 = && \left(\tilde{g}_u \ast g_\ell \ast s_\ell\right)(kT) \label{eq:discrete_continuous_last}
\end{IEEEeqnarray}
where the steps to \Eq{eq:discrete_continuous_second} and \Eq{eq:discrete_continuous_second_to_last} follows from the definitions in \Eq{eq:lmmse_filter} and
\Eq{eq:continuos_to_discrete_time_control_signals} respectively.
Additionally,
\begin{IEEEeqnarray}{rCl}
    \IEEEeqnarraymulticol{3}{l}{ 
        \left(\tilde{g}_u \ast g_0 \ast s_0\right)(kT)} \nonumber\\* \quad
                & = & \left( \tilde{g}_u \ast \left( x_N - (g_u \ast u)\right) \right)(k T) - \sum_{\ell = 1}^N (\breve{h}_\ell \ast s_\ell)[k] \label{eq:discrete_signal_decomposition}
\end{IEEEeqnarray}
where \Eq{eq:discrete_signal_decomposition} follows from \Eq{eq:magic_step} and reversing the steps in \Eq{eq:discrete_continuous_first}-\Eq{eq:discrete_continuous_last}
for $(\tilde{g}_u \ast g_\ell \ast s_\ell)(kT)$.
Replacing $(h_0 \ast s_0)[k]$ with \Eq{eq:discrete_signal_decomposition} in \Eq{eq:estimate} and rearranging results in \Eq{eq:estimate_decomposition}.

\subsection{The Wiener Filter Solution}\label{app:wiener_filter}

Assuming $\hat{s}[.] \eqdef (h_0 \ast s_0)[.]$, $\vct{s}[.]$, 
and $w[k] \eqdef -(\tilde{g}_u \ast x_N)(kT)$ 
to be zero-mean weakly stationary processes and $w[.]$ to be jointly independent of $\vct{s}[.]$,
the minimizing solution to the cost function in \Eq{eq:LMMSE}, i.e., minimizing $\EE{\left|\hat{s}[.] + (\vct{h}^\T \ast \vct{s})[.]\right|^2}$, 
with respect to $\vct{h}[.]$, see \Eq{eq:s_vector} and \Eq{eq:vector_h},
follows by the orthogonality principle as
\begin{IEEEeqnarray}{rCl}%
    (\vct{h}^\T \ast \mat{R}_{\vct{s}})[.] & = & \mat{R}_{\hat{s} \vct{s}}[.], \labell{eq:wiener_equations}%
\end{IEEEeqnarray}
where $\mat{R}_{\vct{s}}[.] \eqdef \EE{(\vct{s} \ast \vct{s}_c^\T)[.]}$
and $\vct{s}_c[.] \eqdef \vct{s}[-.]$.
Convolving both the left hand and right hand side of \Eq{eq:magic_step} with $\tilde{g}_u(.)$, 
enforcing $u(.) = 0$, reversing the steps from \Eq{eq:discrete_continuous_first}-\Eq{eq:discrete_continuous_last}, 
and finally rearranging, gives $\hat{s}[.] = (\vct{g}^\T \ast \vct{s})[.] + w[.]$
where $\vct{g}[k] \eqdef \begin{pmatrix}(\tilde{g}_u \ast g_1 \ast \theta_1)(kT), \dots, (\tilde{g}_u \ast g_N \ast \theta_N)(kT)\end{pmatrix}^\T$.
Subsequently,
\begin{IEEEeqnarray}{rCl}%
    \vct{R}_{\hat{s} \vct{s}}[.] & \eqdef & \EE{(\hat{s} \ast \vct{s}^\T_c)[.]} = (\vct{g}^\T \ast \mat{R}_{\vct{s}})[.] \label{eq:cross_correlation}
\end{IEEEeqnarray}
and plugging \Eq{eq:cross_correlation} into \Eq{eq:wiener_equations}, and assuming a full rank $\mat{R}_{\vct{s}}$, 
results in $h_\ell[.]$ as given by $\breve{h}_\ell[.]$ in \Eq{eq:reference_filter}.

\newpage

\newcommand{\norcas}{IEEE Nordic Circuits and Syst. Conf. (NorCAS)}
\newcommand{\mwscas}{IEEE Int. Midwest Symp. Circuits and Syst. (MWSCAS)}
\newcommand{\iscas}{Proc. IEEE Int. Symp. Circuits Syst. (ISCAS)}
\newcommand{\isscc}{IEEE Int. Solid-State Circuits Conf. (ISSCC)}
\newcommand{\ita}{Information Theory \& Applications Workshop (ITA)}
\newcommand{\asscc}{IEEE Asian Conf. Solid-State Circuits (ASSCC)}

\newcommand{\tcasi}{IEEE Trans. Circuits Syst. I: Reg. Papers}
\newcommand{\tcasii}{IEEE Trans. Circuits Syst. II: Exp. Briefs}
\newcommand{\procIEEE}{Proceedings of the IEEE}
\newcommand{\vlsi}{IEEE Trans. Very Large Scale Integration (VLSI) Systems}

\newcommand{\jssc}{IEEE J. Solid-State Circuits}
\newcommand{\jestcs}{IEEE J. Emerg. Sel. Topics Circuits Syst.}

\newcommand{\phd}[3]{#1, \enquote{#2,} Ph.D. dissertation, #3.}


\begin{thebibliography}{1}


\bibitem{LBWB:11}
H.-A.~Loeliger, L.~Bolliger, G.~Wilckens, and J.~Biveroni,
``Analog-to-digital conversion using unstable filters,''
in \textit{\ita},
La Jolla, CA, USA, Feb. 2011, pp. 1--4,
doi: 10.1109/ITA.2011.5743620.

\bibitem{LW:15} 
H.-A.~Loeliger and G.~Wilckens,
``Control-based analog-to-digital conversion without sampling and quantization,'' 
in \textit{\ita}, 
San Diego, CA, USA, Feb. 2015, pp. 119--122,
doi: 10.1109/ITA.2015.7308975.

\bibitem{M:20}
H. Malmberg, 
``Control-Bounded Converters,'' 
Ph.D. dissertation no. 27025, 
Dept. of Inf. Tech. and Elect. Eng.,
ETH Zürich,
Zürich, Switzerland, 2020,
doi: 10.3929/ethz-b-000469192.

\bibitem{MWL:21}
H. Malmberg, G. Wilckens, and H.-A. Loeliger,
``Control-bounded analog-to-digital conversion,'' 
\textit{Circuits, Syst. Signal Process.},
vol. 41, no. 3, pp. 1223--1254, Mar. 2022,
doi: 10.1007/s00034-021-01837-z.


\bibitem{Schreier:2017}
S.~Pavan, R.~Schreier, and G.~C.~Temes, 
``Continuous-time delta-sigma modulation,''
in \textit{Understanding Delta-Sigma Data Converters,} 
2nd ed., 
Piscataway, NJ: Wiley-IEEE Press, 
2017, ch. 8, pp. 223--258.


\bibitem{R:2011}
J.~M.~de~la~Rosa, 
``Sigma-Delta Modulators: Tutorial Overview, Design Guide, and State-of-the-Art Survey,'' 
\textit{\tcasi},
vol. 58, no. 1, pp. 1--21, Jan. 2011,
doi: 10.1109/TCSI.2010.2097652.


\bibitem{OG:2006}
M.~Ortmanns, F.~Gerfers,
``Continuous-time $\Sigma\Delta$ modulators,''
in \textit{Continuous-Time Sigma-Delta A/D Conversion}, 
Berlin, Germany: Springer Berlin, 2006, ch. 3, pp. 39--84.


\bibitem{FMWLY:2022} 
F.~Feyling, H.~Malmberg, C.~Wulff, H.-A.~Loeliger, and T.~Ytterdal, 
``High-level comparison of control-bounded A/D converters and continuous-time sigma-delta modulators,'' 
in \textit{\norcas}, 
Oslo, Norway, Oct. 2022, pp. 1--5,
doi: 10.1109/NorCAS57515.2022.9934426.


\bibitem{FMWLY:2023} 
F.~Feyling, H.~Malmberg, C.~Wulff, H.-A.~Loeliger and T.~Ytterdal, 
``Design and Analysis of the Leapfrog Control-Bounded A/D Converter,'' 
\textit{\vlsi}, 
pp. 1--10, 2023,
doi: 10.1109/TVLSI.2023.3320279.


\bibitem{Haykin:2002} 
S. Haykin, 
``Least-Mean-Square Adaptive Filters,'',
in \textit{Adaptive filter theory}, 
4th ed. 
Upper Saddle River, NJ: Prentice Hall, 
2002, ch. 5, pp. 231--237.

\bibitem{LDHKPK:2007}
H.-A.~Loeliger, J.~Dauwels, J.~Hu, S.~Korl, L.~Ping, F.R.~Kschischang, 
``The factor graph approach to model-based signal processing,'' 
\textit{\procIEEE},
vol. 95, no. 6, pp. 1295--1322, Jun. 2007,
doi: 10.1109/JPROC.2007.896497.

\bibitem{RACGLM:2006}
J.~Ramirez-Angulo, R.~G.~Carvajal, J.~A.~Galan, and A.~Lopez-Martin,
``A free but efficient class AB two-stage operational amplifier,'' 
in \textit{\iscas},
Island of Kos, Greece, May 2006, pp. 2841--2844,
doi: 10.1109/ISCAS.2006.1693216. 

\bibitem{MAPM:2008}
M.~Miyahara, Y.~Asada, D.~Paik, and A.~Matsuzawa,
``A low-noise self-calibrating dynamic comparator for high-speed ADCs,'' 
in \textit{\asscc}, 
Fukuoka, Japan, Nov. 2008, pp. 269--272,
doi: 10.1109/ASSCC.2008.4708780.


\end{thebibliography}
\end{document}